# A limiting rule for the variability of coding sequences length in microbial genomes

Vasile V. Morariu
*Department of Molecular and Biomolecular Physics*
*National Institute of R&D for Isotopic and Molecular Technology*
*400293 Cluj-Napoca, P.O.Box 700, e-mail: vvm@itim-cj.ro*

May 8, 2008

**Abstract**

The mean length and the variability of coding sequences for 48 genomes of bacteria and archaea were analyzed. It was found that the plotted data can be described by an angular area. This suggests the followings: a) The variability of a genome increases as the mean length increases; b) There is an upper and a lower limit for variability for a given mean length; c) Extrapolation of the upper and lower limits to lower mean values converges to a single point which might be assimilated to a "primordial" cell.. The whole picture is reminding of a process which starts from a single cell and evolves into more and more species which, in turn, show more and more variability.

## 1. Introduction

The microbial genomes consist mainly of coding sequences (CDS) while non-coding sequences represent a minor proportion of the genome. Consequently the composition of coding genes, comprise most of the genome (~90%) and the bacterial genome size shows a strong positive relationship with gene number. The length of a CDS, measured by their number of base pairs, ranges between several hundreds and several thousands. A typical species of bacteria, like *Escherichia coli* contains around $5x10^6$ bases which are organized in several thousands of genes each of them containing a CDS. Such series are quite suitable for the statistical analysis. It should be mentioned that while the correlation properties of DNA at the level of bases have been intensively studied for more than a decade, the series of coding sequence (CDS) lengths received very little attention [1-4]. Although correlation characterization of CDS length series has been previously attempted on microbial genomes, a full statistical characterization has not been reported to date. While the older papers on the subject suggested a weak, long–range correlation the more recent papers, at contrary, brought evidence for short-range correlation and generally a non-uniform organization [3-4]. The present work is part of a project which deals with microbial genome data as a space fluctuating series. Such a series can be characterized from a statistical point of view by their correlation and distribution properties. The subject of this paper is however limited to the variability of the CDS length series in the microbial genomes.

## 2. Materials and Methods

The following 48 genomes of bacteria and archaea were investigated: *Acidovorax avenae subspecie citrulli* AAC00-*1, Anabaena variabilis* ATCC 29413*, Aeropyrum pernix* K1*, Archaeoglobus fulgidus, Azorhizobium caulinodans* ORS571*, Bacilus subtilis, Bacillus pumilis* SAFR-032*, Bacillus halodurans* C-125 *Bacillus cereus* TCC14*, Bacillus thuringiensis serovar konkukian* str. 97-27*, Bacteroides thetaiotaomicron* VPI-548*, Bradyrhizobium japonicum* USDA 110*, Clostridium beijerinckii* NCIMB 8052*, Delftia acidovorans* SPH-1*, Escherichia coli* APEC 01*, E.coli* O157:H7 strain Sakai*, E.coli* O157:H7 EDL 933*, E.coli* CFT073*, E.coli* K12*, Enterococus faecalis* V583*, Flavobacterium johnsoniae* UW101*, Frankia sp.EAN1pec, Hahella cheujuensis* KCTC 2396*, Haemophilus influenzae* strain 86-028 NP*, Haemophilus influenzae* strain ATCC51907/KW20/Rd*, Helicobacter pylori* 266*, Lactobacillus plantarum* strain WCFS1*, Lactobacillus casei* ATCC 334*, Mycoplasma penetrans* HF-2*, Mycoplasma pneumoniae* M129*, Mycobacterium smegmatis* strain MC2 155*, Mycrocystis aeruginosa* NIES-843*, Methanococcoides burtonii* DSM 6242*, Nocardia farcinica* IFM 10152*, Pseudomonas entomophila* strain L48*, Photorhabdus luminiscens subspeciae laumondii TTO1, Rhodococcus* sp. RHA1*, Sulfolobus solfataricus, Sorangium cellulosum* 'Soce 56'*, Staphilococcus aureus subspeciae aureus JH1, S. aureus subsp. aureus* JH9*, S. aureus subsp. aureus* Mu3*, S. aureus subsp. aureus* Mu50*, S.aureus subsp. aureus* MW2*, S. aureus subsp. aureus* N315*, S. aureus subsp. aureus* NCTC 8315*, S. aureus subsp. aureus strain Newman, S. aureus subsp. aureus* USA 300 TCH 1516*, Xanthobacter autotrophicus Py2.* from the EMBL-EBI data base. The species were selected such as to cover all main divisions of microbes as well as some of the main models used in literature (*E. coli* and *B. subtilis*). In case of *E. coli* and *Staphilococus aureus* various strains were included in order to compare the variability of the statistical properties of strains.

The extraction of the data, from the EMBL-EBI data base, was done with a program written in MATLAB. The length of coding sequences, expressed as number of base pairs, was calculated as the difference between the start and the end position of CDS in the genome. The mean value of the length is also available in the proteome section of EMBL-EBI data base. Variability of the CDS length series can be estimated by calculating the standard deviation of the series.

## 3. Results and Discussion

The variability of the length of coding sequences in a genome is illustrated in figure 1. The plot shows that the lengths vary to a considerable extent along the genome.

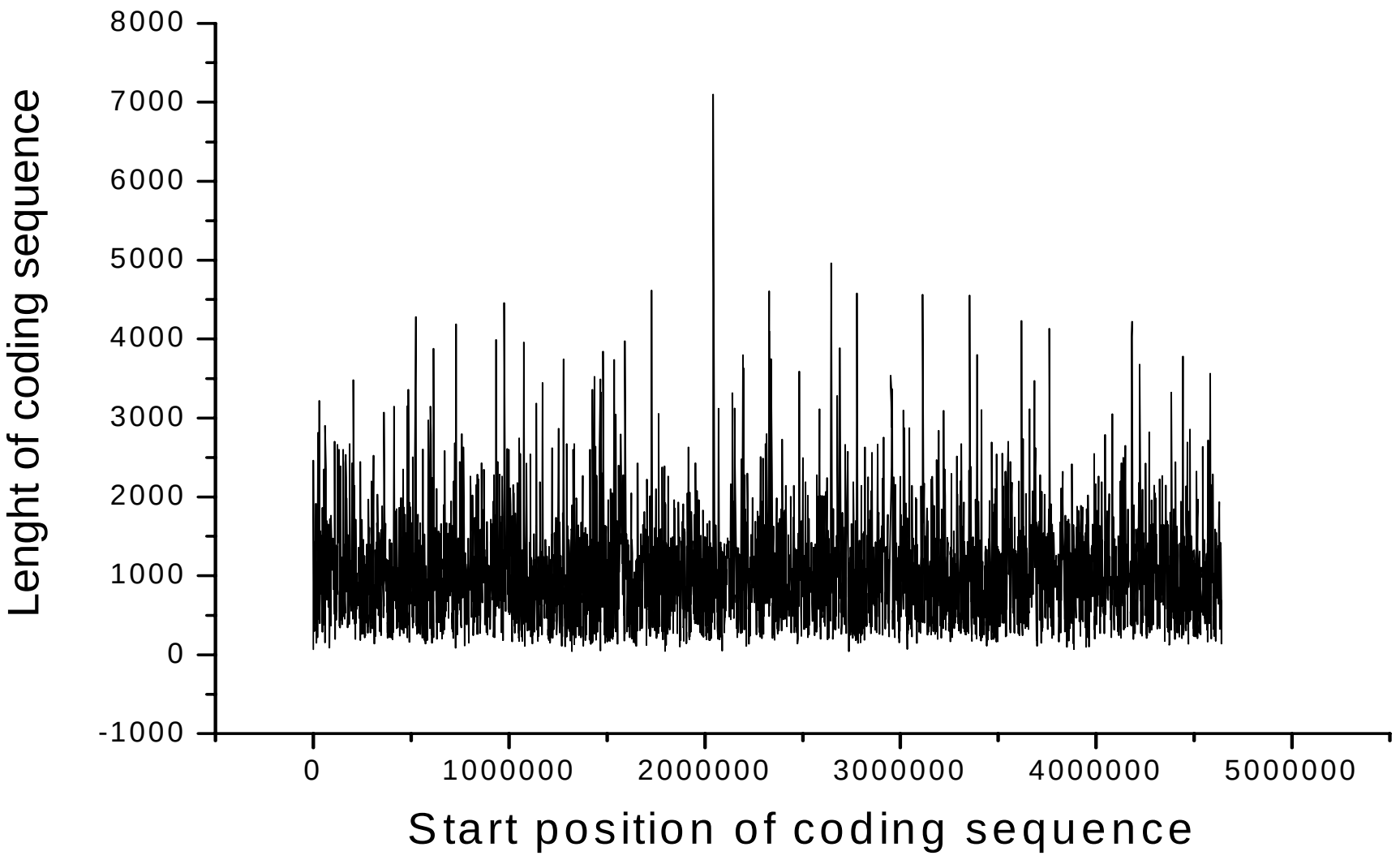

Figure 1. The length of coding sequences (bp) in the genome of *Escherichia coli* K12 strain. The plots illustrate the diversity of the coding sequence lengths along the genome.

As the mean value of CDS length increases variability of the series also increases (figure 2). The interesting aspect of this plot is twofold: a) It shows there is a lower and an upper limit for the variability of CDS lengths. These limits seem quite linear and they tell that microbial species below and above these limits cannot exist; b) As the linear limits delimitate an angular space the lower the mean value of the CDS length the lower the variability which may exist for similar mean values. This points out to a hypothetic organism which has the lowest possible mean value and variability at around x=550 bp and y=250 bp. The whole picture looks like an evolving image. Starting from a single most simple species the picture evolves into more and more diversified species.

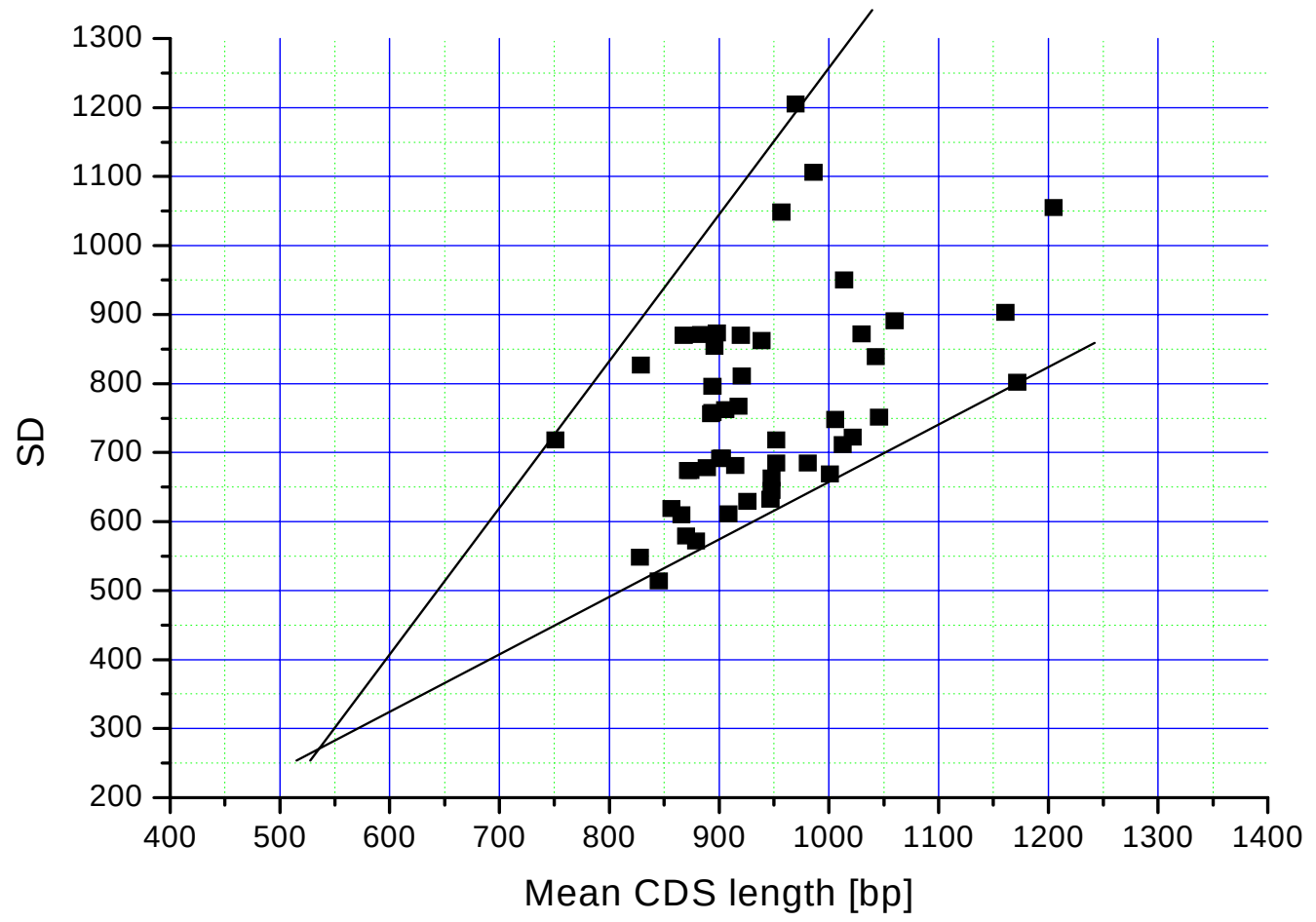

Figure 2. Variability (measured as standard deviation) of the coding sequence lengths *versus* the mean length for 48 bacterial and archaeal genomes. Note the angular space for the distribution of data and the intersection point of the two extrapolated limits.

The lower limit of the angular space is illustrated in figure 3. It also includes the identification of each microbial species. This lower limit is remarkably linear and includes species belonging to quite different divisions. They include archaea, and bacteria (Divisions: bacteroidetes, firmicutes, and proteobacteria). Archaea are located in the lower part of the plot, being characterized by lower mean lengths and SD values.

The problem of bacteria evolution is under continuous debate and figure 2 should not necessary be regarded as a time like picture (where times flow to the right). On the other hand is very tempting to look at it as a time evolution as the “origin” of the angular area might suggests a common ancestor for all microbes. If this is a distant ancestor then it is unlikely to find real microbes with close characteristics to “origin”. A better answer should be postponed until more data will fill the plot in figure 2. However it suggests to direct investigation to mean lengths around 800 bp and lower. Also it is quite possible that the 48 species analyzed in this work cannot give a precise image about the lower as well as the upper limits. Consequently future work will need to analyze more microbial species in order to strengthen the upper and lower limits and to define more precisely the "primordial" species. It will also be very interesting to find how close to the "primordial" cell the real microbes may be located.

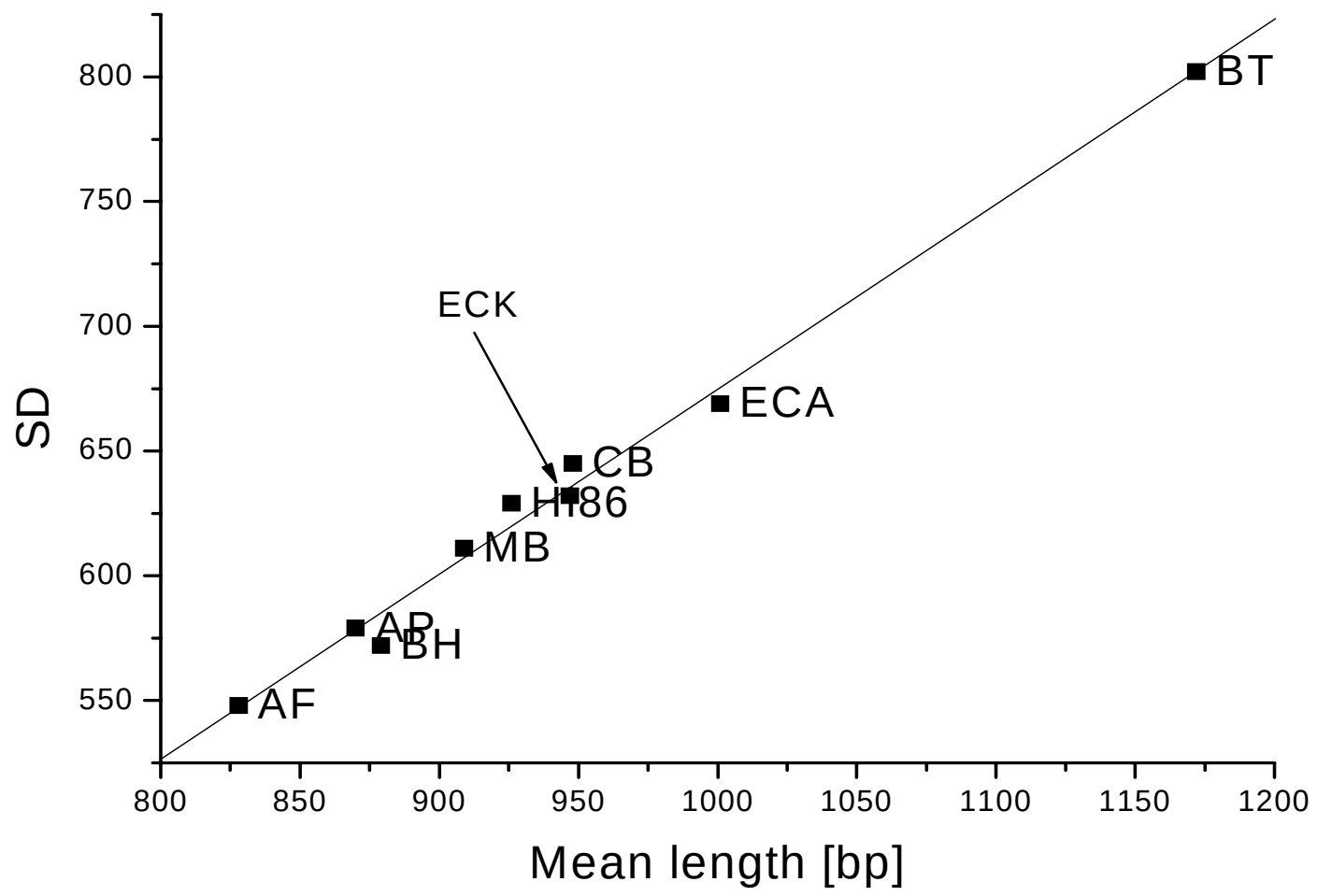

Figure 3. The lower limit of the plot illustrated in figure 2. The microbial species are: AF- *Archaeoglobus fulgidus* DSM 4304; AP- *Aeropyrum pernix* K1; BH- *Bacillus halonduras* C-125; MB- *Methanococcoides burtonii* DSM 6242; HI86- *Haemophilus influenzae* 86-028NP; ECK- *Escherichia coli* K12; CB- *Clostridium beijerrinckii* NCIMB 8052; ECA- *Escherichia coli* APEC O1; BT- *Bacteroides thetaiotaomicron* VPI5482.

## Acknowledgements

This work was supported by a project financed by the Romanian Authority for Scientific Research. Thanks are due to Professor Octavian Popescu.